\documentclass[twocolumn, 10pt]{article}

\usepackage[hidelinks]{hyperref}

\def\equationautorefname~#1\null{equation (#1)\null}

\usepackage{graphicx}
\usepackage[separate-uncertainty]{siunitx}
\usepackage{xcolor}
\usepackage{amsmath}
\usepackage{bm}
\usepackage{gensymb}

\usepackage{epstopdf}
\usepackage[sort&compress,numbers,square]{natbib}	%

\usepackage[margin=1.8cm]{geometry}

\newcommand{\mean}[1]{\overline{#1}}
\newcommand{\dtens}[1]{\bm{D}[#1]}
\newcommand{\nm}{\,\si{\nano\meter}}
\newcommand{\um}{\,\si{\micro\meter}}
\newcommand{\mm}{\,\si{\milli\meter}}
\newcommand{\cm}{\,\si{\centi\meter}}

\newcommand{\mL}{\,\si{\milli\liter}}
\newcommand{\uL}{\,\si{\micro\liter}}

\newcommand{\mM}{\,\si{\milli\textsc{m}}}

\usepackage{authblk} 

\title{Flexibility-induced effects in the Brownian motion of colloidal trimers}

\author[1,*]{Ruben W. Verweij}
\author[2,3,*]{Pepijn G. Moerman}
\author[2]{Nathalie E. G. Ligthart}
\author[1]{Loes P. P. Huijnen}
\author[2,4]{Jan Groenewold}
\author[2]{Willem K. Kegel}
\author[3]{Alfons van Blaaderen}
\author[1]{Daniela J. Kraft}

\affil[*]{These authors contributed equally to this work}
\affil[1]{\small Huygens-Kamerlingh Onnes Lab, Universiteit Leiden}
\affil[2]{\small Debye Institute for Nanomaterials research, Department of Chemistry, Utrecht University}
\affil[3]{\small Debye Institute for Nanomaterials research, Department of Physics, Utrecht University}
\affil[4]{\small Academy of Advanced Optoelectronics, South China Normal University, Guangzhou}
\affil[ ]{Corresponding email: \url{kraft@physics.leidenuniv.nl}, \url{A.vanBlaaderen@uu.nl}, \url{W.K.Kegel@uu.nl}}

\date{}

\begin{document}

\twocolumn[
  \begin{@twocolumnfalse}
    \vspace*{-1.0cm}
    \maketitle
    \vspace*{-1.0cm}
    \begin{abstract}%
    Shape changes resulting from segmental flexibility are ubiquitous in
    molecular and biological systems, and are expected to affect both the diffusive motion and (biological)
    function of dispersed objects. The recent development of colloidal structures with freely-jointed bonds have now made a direct experimental investigation of diffusive shape-changing objects possible. Here, we show the effect of segmental
    flexibility on the simplest possible model system, a freely-jointed cluster
    of three spherical particles, and validate long-standing theoretical predictions. We find that in addition to the rotational diffusion time, an analogous conformational diffusion time governs the relaxation of the
    diffusive motion, unique to flexible assemblies, and that their translational diffusivity differs by a small but measurable amount. We also uncovered a Brownian quasiscallop mode, where diffusive motion is coupled to Brownian shape changes. Our findings could have implications for molecular and biological systems where diffusion plays an important role, such as functional site availability in lock-and-key protein interactions.
\end{abstract}
    \vspace*{0.8em}
  \end{@twocolumnfalse}
]

\begin{figure}[ht]
    \centering
    \includegraphics[scale=1]{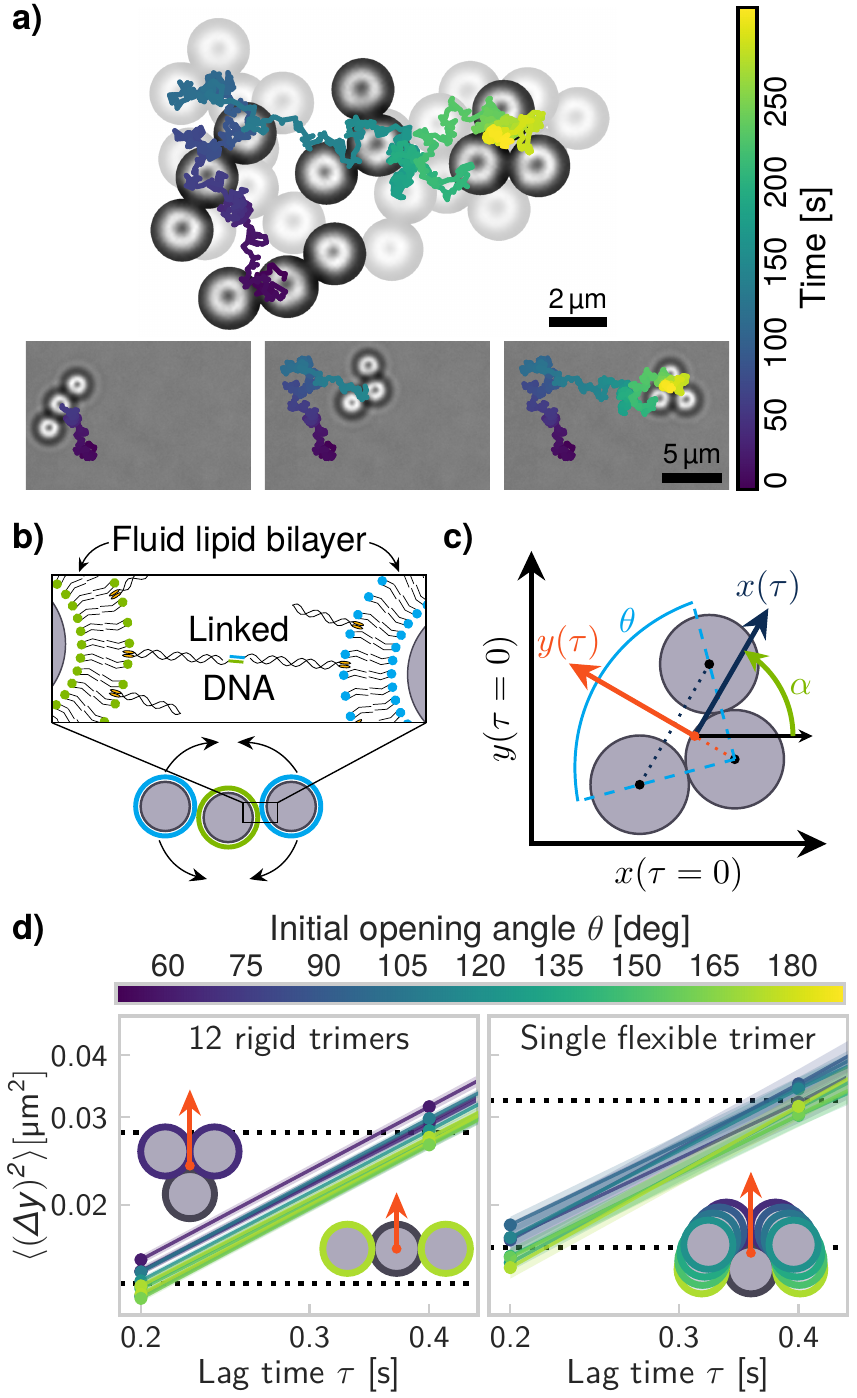}

    \caption{\textbf{Diffusion of flexible trimers} \textbf{a)} Overlay of
        bright-field microscopy images of a flexible trimer with the position
        of its center of mass as function of time. \textbf{b)} Schematic (not
        to scale) of flexible trimers that are self-assembled from colloid
        supported lipid bilayers. We inserted DNA linkers into the fluid lipid
        bilayer surrounding the particle, resulting in bonded particles that
        can rearrange with respect to each other. \textbf{c)} Illustration of
        the body-centered coordinate system. \textbf{d)} The mean squared
        displacement of rigid and flexible trimers. The translational mean
        squared displacement of flexible trimers in the $y$-direction is angle
        dependent for short lag times, at longer lag times this angle
        dependence is no longer present due to rotational and conformational
        relaxation, which happens on a shorter timescale than for rigid trimers
    (raw data, not scaled with friction coefficients).\label{fig:fig1}} 

\end{figure}

\section{Introduction} 

Many (macro)molecular systems display segmental flexibility, e.g.\ bio-polymers
such as transfer RNA \cite{Mellado1988}, intrinsically disordered proteins
\cite{Ishino2014}, myosin \cite{Mellado1988}, immunoglobulins
\cite{Mellado1988}, and other antibodies
\cite{Yguerabide1970,Torre1994,Burton1987,Barbato1992}.  For most of these
systems, the flexibility not only affects the motion of the complex but also
its (biological)
function~\cite{Serdyuk2007,Torre1994,Gregory1987,Yguerabide1970,Illien2017}.
For example, proteins often function through shape-dependent lock-and-key
interactions where active sites of enzymes are reshaped during the interaction,
leading to an induced fit~\cite{Koshland1995}. Additionally, enzymes like
adenylate kinase can accelerate biochemical reactions with remarkable
specificity and efficacy thanks to a flexible ``lid" that opens and closes at
each reaction cycle.  Because shape has a large effect on the diffusive motion
of structures at the short timescales relevant to these reactions, it is
expected that the diffusion of re-configurable objects is different from rigid
ones~\cite{Mellado1988,Nagasaka1985,Fixman1981,Akcasu1982}. Moreover,
Adeleke-Larodo~et~al.\ recently proposed~\cite{Adeleke2019} that changes in an
enzymes flexibility upon substrate binding could be responsible for the
observed enhanced diffusion of active enzymes~\cite{Riedel2015, Sengupta2013}.
Therefore, a rigorous understanding of enzyme function and diffusion requires
quantitative knowledge of protein flexibility \cite{Rundqvist2009}.

However, direct experimental measurements of flexibility in molecular systems
are challenging because they require single-molecule measurement techniques
with high spatial and temporal resolution. One way to circumvent this problem
is to employ colloidal particles, which have been used as model systems for
(macro)molecular structures~\cite{Einstein1905,Sutherland1905,Perrin1909},
because of their unique combination of microscopic size and sensitivity to
thermal fluctuations. Studies on the Brownian motion of rigid colloids of
various shapes such as ellipsoids~\cite{Han2006,Meunier1994,Zahn1994},
boomerangs~\cite{Chakrabarty2014,Chakrabarty2016,Koens2017}, and
clusters~\cite{Kraft2013,Fung2013} have revealed that shape affects the
diffusive motion at short timescales. Additionally, displacements are larger in
directions that correspond to smaller hydrodynamic
drag~\cite{Han2006,Kraft2013,Fung2013,Chakrabarty2014,Chakrabarty2016,Hoffmann2009}
and different diffusive modes can be coupled, e.g.\ helical particles rotate as
they translate and vice-versa~\cite{Butenko2012}. At longer timescales, the
influence of particle shape decreases because of rotational
diffusion~\cite{Han2006}.

While rigid assemblies have been extensively studied, little is known about the
effect of flexibility. In order to numerically and experimentally investigate
the effect of segmental flexibility, we study a simple model system consisting
of a freely-jointed chain of three spherical colloidal particles, called
flexible trimers or ``trumbbells'' \cite{Harvey1983,Roitman2005}.  Numerical
models were proposed to capture the diffusion of segmentally flexible
objects~\cite{Wegener1982,Harvey1983,Wegener1985} and the long time diffusive
motion was predicted to be determined by the shape average of the instantaneous
diffusivities (so-called rigid-body
approximation)~\cite{Torre1994,HydroSub,Iniesta1988}.  For the first time, we
are able to test these models using direct experimental measurements of the
diffusion of colloidal particles, thanks to the recent development of colloidal
structures with freely-jointed
bonds~\cite{VanderMeulen2013,VanderMeulen2015,Chakraborty2017,Zhang2017,Zhang2018,Mcmullen2018,Rinaldin2019},
and flexible chains~\cite{Vutukuri2012}. First, we discuss the short-time
diffusion tensor of the flexible trimers, which we compared to numerical
calculations and found a good agreement. Furthermore, we uncovered a Brownian
quasiscallop mode, where diffusive motion is coupled to Brownian shape
changes. Next, we considered the diffusive behaviour at longer timescales and
found that in adition to the rotational diffusion time, an analogous
conformational diffusion time governs the relaxation of the diffusive motion,
unique to flexible assemblies.

\section{Methods}

\subsection{Experimental} 

Flexible clusters of three colloidal supported lipid bilayers (CSLBs) were
prepared as described in previous work
\cite{VanderMeulen2013,VanderMeulen2015,Chakraborty2017,Rinaldin2019}.  To test
the generality of the results presented here, we used two particle sizes,
namely \SIlist{1.93;2.12}{\um} silica particles, with different methods of
functionalization.

The CSLBs consisting of \SI{2.12}{\um} silica particles were prepared as
described in our recent work \cite{Rinaldin2019}. Briefly, the particles were
coated with a fluid lipid bilayer by deposition of small unilamellar vesicles
consisting of \SI{98.8}{\mole\percent} DOPC (($\Delta$9-Cis)
1,2-di\-o\-le\-oyl-sn-gly\-ce\-ro-3-phos\-pho\-cho\-line),
\SI{1}{\mole\percent} DOPE-PEG(2000)
(1,2-di\-o\-le\-oyl-sn-gly\-ce\-ro-3-phos\-pho\-e\-tha\-nol\-a\-mine-N-[me\-thox\-y(po\-ly\-e\-thy\-lene
gly\-col)-2000]) and \SI{0.2}{\mole\percent} TopFluor-Cholesterol
(3-(di\-pyr\-ro\-me\-thene\-bo\-ron
di\-flu\-o\-ride)-24-nor\-cho\-les\-te\-rol) or DOPE-Rhodamine
(1,2-di\-o\-le\-oyl-sn-gly\-ce\-ro-3-phos\-pho\-e\-tha\-nol\-a\-mine-N-(lis\-sa\-mine\-rho\-da\-mine
B sulf\-o\-nyl)). The bilayer coating was performed in a buffer at pH 7.4
containing \SI{50}{\mM} sodium chloride (NaCl) and \SI{10}{\mM}
4-(2-Hy\-dro\-xy\-e\-thyl)-1-pi\-per\-a\-zine\-e\-thane\-sul\-fo\-nic acid
(HEPES). We added double-stranded DNA (of respectively strands DS-H-A and
DS-H-B, see the Supplementary Information) with an 11 base pair long sticky end
and a double stearyl anchor, which inserts itself into the bilayer via
hydrophobic interactions (see \autoref{fig:fig1}b). When two particles with
complementary DNA linkers come into contact, the sticky ends hybridize and a
bond is formed.  Self-assembly experiments were performed in a different buffer
of pH 7.4, containing \SI{200}{\mM} NaCl and \SI{10}{\mM} HEPES. We imaged 21
trimers of \SI{2.12}{\um} CSLBs, that were formed by self-assembly in a sample
holder made of po\-ly\-a\-cryl\-a\-mide (PAA) coated cover glass. The PAA
functionalisation was carried out using a protocol \cite{VanDerWel2016} which
we modified by adding \SI{0.008}{\mole\percent} bis-a\-cryl\-a\-mide and
performing the coating under a nitrogen atmosphere, both of which resulted in a
more stable coating. Using an optical microscope, we imaged the clusters for
\SI{5}{\minute} at frame rates between \SIrange{5}{10}{fps}.  Particle
positions were tracked using a custom algorithm \cite{Rinaldin2019} available
in TrackPy by using the \texttt{locate\_brightfield\_ring} function
\cite{TrackPy}.

Additionally, we analysed 9 trimers of \SI{1.93}{\um} CSLBs, with silica
particles purchased from Microparticles GmbH (product code SiO$_2-$R-B1072).
For these particles, we used a similar protocol to form supported lipid
bilayers with only 2 minor modifications: first, the lipid composition we used
was \SI{98.9}{\mole\percent} DOPC, \SI{1}{\mole\percent} DOPE-PEG(2000) and
\SI{0.1}{\mole\percent} DOPE-Rhodamine. Second, we added Cy3-labeled DNA with a
self-complementary 12 base pair sticky end and a cholesterol anchor that
inserts itself into the lipid bilayer due to hydrophobic interactions. We used
the DNA sequence from Leunissen \textit{et al}. \cite{Leunissen2009} (see
Supplementary Material, strands PA-A and PA-B).

To image the \SI{1.93}{\um} CSLBs we used a flow cell produced as detailed in
the Supplementary of Montanarella \textit{et al}.\cite{Montanarella2018} As the
base of our flow cell we used a single capillary with dimensions
$3~\cm~\times~2~\mm~\times~200~\um$. To prevent the lipid coated clusters from
sticking to the class capillary, we coated the inside of the capillary with
poly(2-hydroxyethyl acrylate) (pHEA) polymers. To this end, we first flushed the cell
with consecutively $2~\mL$ $2~\mM$ NaOH solution, $2~\mL$ water and $2~\mL$
EtOH. We then functionalized the glass surface with the silane coupling agent
3-(methoxysilyl)propyl methacrylate (TPM) by filling the flow cell with a
mixture of $1~\mL$ EtOH, $25~\uL$ TPM, and $5~\uL$ $25~vol\%$~NH$_3$ in water
and leaving it for $1$ hour. We then washed and dried the flow cell by flushing
with $2~\mL$ ethanol and subsequently with Nitrogen. We grew pHEA brushes from
the surface through a radical polymerization by filling the cell with a mixture
of $2.5~\mL$ EtOH, $500~\uL$ HEA and $20~\uL$ Darocur 1173 photoinitiator. We
initiated the reaction by placing the cell under a UV lamp $~\lambda=360~\nm$
for $10$~minutes. Finally, we flushed the cells with $10~\mL$ EtOH or Millipore
filtered water. We stored the coated cells filled with EtOH or Millipore
filtered water and for no more than $1$ day. Self-assembly experiments were
performed in a  buffer of pH 7.4, containing \SI{50}{\mM} NaCl and \SI{10}{\mM}
HEPES. We imaged 9 freely-jointed trimers and 13 rigid trimers stuck in various
opening angles shown in the Supplementary Information for 30 minutes with a
frame rate of 5 frames per second. Particle positions were tracked using the
2007 Matlab implementation by Blair and Dufresne of the Crocker and Grier
tracking code. \cite{Crocker1996}
    
\subsection{Diffusion analysis}

For all analysis, we only selected trimers that showed all bond angles during
the measurement time, experienced no drift and were not stuck to the substrate.
After the particle positions were tracked, we determined the short-time
diffusivity of the trimers as described by \autoref{eq:dtensor}.

The three friction correction factors that account for substrate friction were
determined in the following way:%
\begin{align} \phi_{tt} &=
\left\langle\bm{D}[tt]_t / (\sigma_e \bm{D}[tt]_{e,0})\right\rangle \nonumber
\\ \phi_{(\alpha\alpha,\theta\theta)} &= \left\langle\bm{D}[(\alpha\alpha,
\theta\theta)]_t / (\sigma_e^3
\bm{D}[(\alpha\alpha,\theta\theta)]_{e,0})\right\rangle \nonumber \\ \phi_{ij}
&= \sqrt{\phi_{ii}\phi_{jj}} \quad \textrm{for} \quad i\neq j,\label{eq:phi}
\end{align}%
where $\bm{D}[ij]_k$ denotes the theoretical ($k=t$) or experimental ($k=e$)
diffusion tensor element and $\sigma_e$ the experimental particle radius. The
subscript $tt$ denotes the translational component of the diffusivity. These
factors were determined separately for each experiment, because differences in
surface and particle functionalisations resulted in differences in
substrate-particle and particle-particle friction, that in turn affect the
diffusivity of the cluster. We separated the correction factors into these
three factors because different modes of diffusion are expected to lead to
different amounts of friction with the substrate \cite{Padding2010}.

We calculated the elements of the diffusion tensor given in
\autoref{eq:dtensor} separately for all trimers. For each pair of frames, we
determined the initial average opening angle $\mean{\theta}$ of the trimer
between $t$ and $t+\tau_{short}$, with $\tau_{short}=\SI{0.25}{\second}$. Then,
we stored the diffusion tensor elements separately for each initial opening
angle. For short times up to $\tau_{short}=\SI{0.25}{\second}$, we used a bin
size of \SI{15}{\degree} while for longer times, we used two bins of
\SI{60}{\degree} covering the range of $[\SI{60}{\degree}, \SI{120}{\degree})$
and $[\SI{120}{\degree}, \SI{180}{\degree}]$. We scaled each element with the
friction factors we obtained for that measurement, based on the diffusion
coefficient for lag times up to $\tau_{short}$. The average diffusion tensor
elements were then obtained by fitting the overall slope of the mean (squared)
displacements of all the individual diffusion tensor elements as a function of
lag time (see \autoref{fig:fig3}a, c, e and \autoref{fig:fig4}a, c). We used a
linear function (with zero intercept) divided into ten segments with slopes $2
D_i$ (spaced evenly on a log scale), which correspond to the $i$th diffusion
coefficient for those lag times. This resulted in the average diffusion tensor
for all binned average opening angles $\mean{\theta}$ as a function of the lag
time $\tau$. For fitting, we used a standard least squares method and we
estimated the error using a Bayesian method to find an estimate of the
posterior probability distribution, by using a Markov chain Monte Carlo (MCMC)
approach as implemented in the Python packages lmfit~\cite{lmfit} and
ecmee~\cite{emcee}.  We estimated the autocorrelation time $\tau_{acor}$ of the
chain using the built-in methods and ran the analysis for at least
$100\tau_{acor}$ steps, where we discarded the first $2\tau_{acor}$ steps
(corresponding to a burnin phase) and subsequently used every other
$\tau_{acor}/2$ steps (known as thinning). The reported values correspond to
the maximum likelihood estimate of the resulting MCMC chain, the reported
uncertainties correspond to the minimum and maximum of the obtained posterior
probability distribution.

\subsection{Hydrodynamic modelling} 

The diffusion of segmentally flexible objects can be described using
hydrodynamic modelling \cite{Harvey1983,Diaz1994}.  To compare our experimental
results to these predictions, we followed the procedure described by Harvey and
coworkers. Of the seven degrees of freedom in three dimensions (three
translational, three rotational, one internal degree of freedom)
\cite{Harvey1983}, we considered only the four degrees of freedom of interest
for our quasi-two dimensional system of sedimented clusters. Briefly, following
the method outlined by Harvey and coworkers \cite{Harvey1983}, we determined
the hydrodynamic resistance (or friction) tensor $\bm{R}_{0}$ with respect to
the central particle. Using this resistance tensor, we calculated the diffusion
tensor $\bm{D}_{0} = kT \bm{R}_0^{-1}$, to which we apply the appropriate
coordinate transformation to obtain the $7 \times 7$ diffusion tensor
$\bm{D}_{com}$ relative to the center of mass of the cluster.  We chose the
center of mass as reference point because this is the best approximation of the
center of diffusion of a flexible particle: in fact, it was found to be a
better choice than either the center of diffusion or resistance of a rigid
cluster of the same shape \cite{Harvey1983}.  We have also calculated the
diffusion tensor with respect to the central particle and these results are
shown in the Supplementary Information.  

The diffusivity of flexible colloidal clusters can be modelled using bead or
bead-shell models \cite{Carrasco1999b} and we employed both methods. For the
bead model, we modelled the trimer using three beads (diameter of \SI{2}{\um})
and for the bead-shell model, we modelled the trimer using approximately
\numrange{2500}{9500} smaller beads with bead radii from \SIrange{54}{31}{\nm}
respectively, where the beads where placed to form three \SI{2}{\um} shells. We
followed existing methods \cite{Bloomfield1967,Filson1967} for constructing the
bead shell model: to summarize, the positions of the small beads were
calculated by placing them on concentric circles, starting at the equator of an
individual \SI{2}{\um} sphere and continuing the process towards the poles of
the sphere using circles of decreasing radius and finally putting one sphere at
each of the poles. Three spherical bead-shell models were then put together to
form a trimer and we removed overlapping beads at the contact points between
the particles. Examples of the model are shown in the Supplementary
Information.

Because drag forces act on the surface of the particles, the bead-shell model
is more accurate in describing the diffusive properties of the clusters
\cite{Bloomfield1967,Filson1967,HydroSub}. The accurate consideration of
hydrodynamic effects was found to be important for the segmentally flexible
system we study: hydrodynamic interactions lead to a slower decay of the
auto-correlation of the particle shape \cite{Ermak1978} and lead to an increase
in the translational diffusivity \cite{Mellado1988,Torre1994}. We compare our
experimental data to such a bead-shell model because it describes our
experimental data more accurately than the simple bead model, which is
discussed in the Supplementary Information. 

To calculate the diffusion tensor elements, we used the Rotne-Prager-Yamakawa
(RPY) \cite{Rotne1969,Yamakawa1970} interaction tensor $\bm{T}_{ij}$ to model
hydrodynamic interactions between particles $i$ and $j$:%
\begin{align}
    \bm{T}_{ij} &= \frac{1}{8\pi\eta_0R_{ij}} \left[\bm{I}+\frac{\bm{R}_{ij}\bm{R}_{ij}}{R^2_{ij}} + \frac{2\sigma^2}{R^2_{ij}} \left( \frac{\bm{I}}{3} - \frac{\bm{R}_{ij}\bm{R}_{ij}}{R^2_{ij}} \right) \right],
\label{eq:hydro_interaction_tens}
\end{align}%
where $\sigma$ is the particle radius, $\bm{R}_{ij}$ is the vector between
particles $i$ and $j$, $\bm{I}$ is the $3 \times 3$ identity matrix, $\eta_0$
is the viscosity of the medium. Using the RPY tensor prevents singularities
that may lead to the large, non-physical numerical fluctuations
\cite{Carrasco1999} found when using lower order terms (Oseen tensor), higher
order terms or multi-body effects \cite{Phillies1984}.

We used the RPY tensor to model the hydrodynamic interactions between the beads
and followed the procedure outlined by Harvey and coworkers \cite{Harvey1983}
to obtain the diffusion tensor, as explained in the main text. This was done
for all small bead radii and we used a linear extrapolation to zero bead size
to obtain the final diffusion tensor elements \cite{Bloomfield1967,Filson1967}.
Additionally, we used HydroSub~\cite{HydroSub} to model the diffusivity of
rigid trimers of the same opening angles.

\section{Results and Discussion}

The flexibly linked colloidal trimers are made by self-assembly of colloid
supported lipid bilayers~
\cite{VanderMeulen2013,VanderMeulen2015,Chakraborty2017,Rinaldin2019}.
Briefly, spherical colloidal silica particles are coated with a fluid lipid
bilayer. DNA linkers with complementary sticky ends are inserted into the
bilayer using a hydrophobic anchor. The particles are self-assembled by
hybridization of the DNA sticky ends, which provide strong and specific
interactions. The trimers are freely-jointed because the DNA linkers can
diffuse on the fluid lipid bilayer that surrounds the particles (see
\autoref{fig:fig1}b). The clusters undergo translational and rotational
diffusion while they are also free to change their shape (see
\autoref{fig:fig1}a and Supplementary Movie 1). For simplicity, we used heavy
silica particles so that their mobility is confined to the bottom of the
container by gravity, which leads to two-dimensional Brownian motion.

For rigid objects in two dimensions, the diffusive motion can be described by a $3 \times 3$
diffusion tensor calculated from the linear increase of the mean squared
displacements of the particle as function of lag time~\cite{Happel2012}. For
flexible objects, this diffusion tensor has to be extended with an additional
degree of freedom~\cite{Harvey1983} for each internal deformation mode (here:
one), and we therefore consider the $4 \times 4$ diffusion tensor $\bm{D}[ij]$.
Here, $i, j \in [x, y, \alpha, \theta]$ are elements of a body-centered coordinate
system (see \autoref{fig:fig1}c) at the center of mass. We chose the center of
mass as reference point, because for flexible objects, it is more appropriate
than either the center of diffusion or resistance of a rigid cluster of the
same shape \cite{Harvey1983}. In this coordinate system the $y$-axis is
perpendicular to the end-to-end vector and points away from the central
particle, and the direction of the $x$-axis is chosen to form a right-handed
coordinate system. We label the opening angle of the trimer $\theta$ and the
(anti-clockwise) rotation angle of the $x$-axis with respect to the lab frame $\alpha$. We align the lab frame such that it coincides with the body-centered coordinate system at $\tau=0$.

Shape determines the diffusion tensor for rigid objects and therefore we expect
it to be important for flexible objects as well, but due to its flexibility,
the cluster shape is continuously changing. Therefore, we categorize the
trajectories by their (initial) average opening angle $\mean{\theta}$ of the
smallest lag time interval and we use angular bins to summarize the results.
The short-time diffusion tensor is calculated from experimental measurements in
the following way: %
\begin{align} \dtens{ij}(\mean{\theta}) &\equiv \frac{1}{2}
    \phi_{ij} \frac{\partial\langle\Delta i \Delta j\rangle_\tau}{\partial
    \tau}, \label{eq:dtensor} 
\end{align}%
with $\tau$ the lag time between frames, $\langle \cdots \rangle_\tau$ denotes
a time average over all pairs of frames $\tau$ apart and $\Delta i = i(t+\tau)
- i(t)$, $\phi_{ij}$ is a correction factor that accounts for particle-particle
and particle-substrate friction (see Methods section). The correction factors $\phi_{ij}$ are a first-order approximation to model the wall effect of the glass surface, that for translational diffusion agrees closely with predictions from hydrodynamic theory (see the Supplementary Information). We evaluated equation~\ref{eq:dtensor} at $\tau = 0.25s$, set by the frame rate of our camera.

\begin{figure*}[ht]
    \centering
    \includegraphics[scale=1]{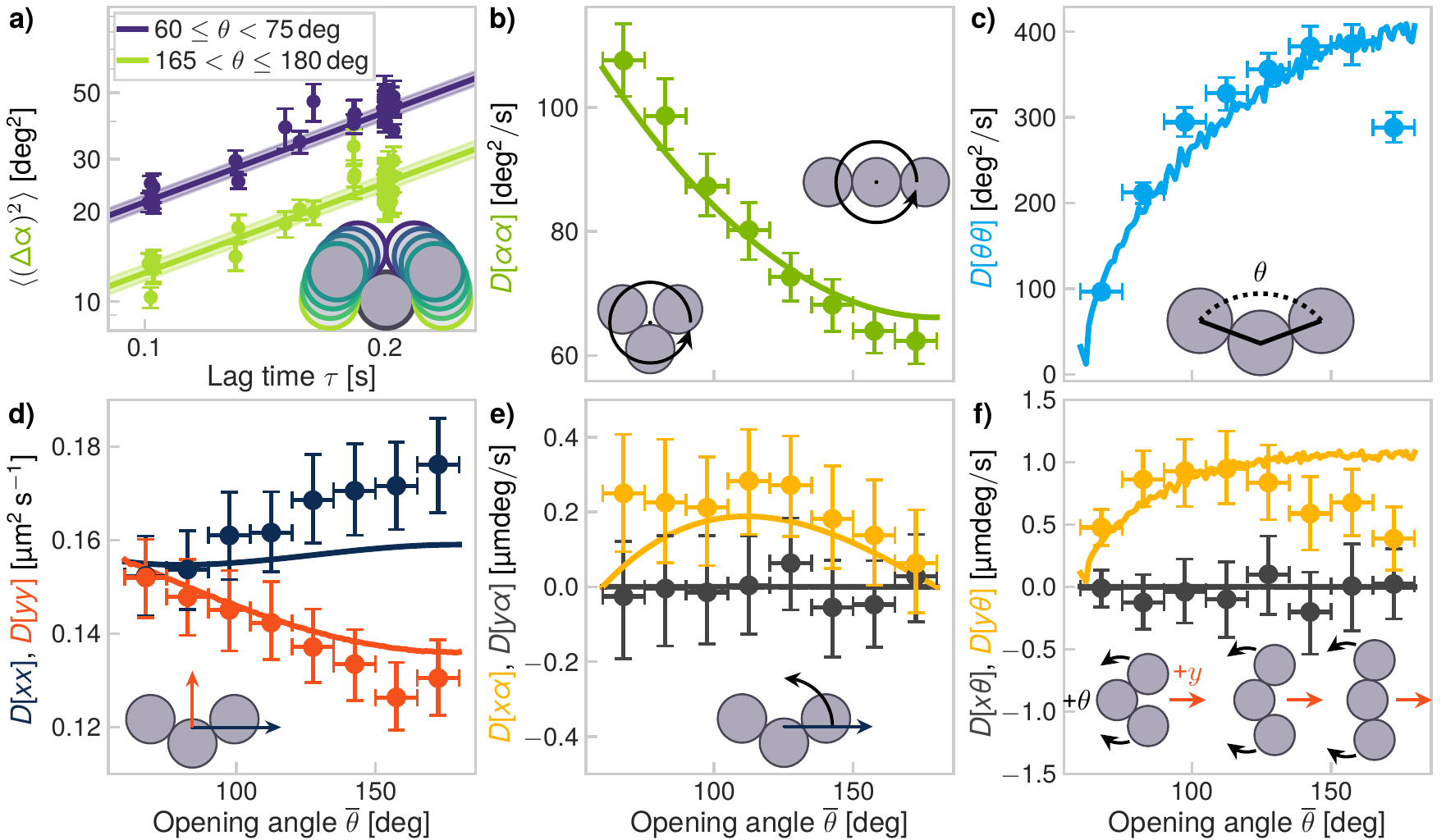}

    \caption{\textbf{Short-time translational, rotational, internal and coupled
        diffusivity} of flexible trimers (up to \SI{0.25}{\second}). \textbf{a)}
        Mean squared rotational displacements for lag times up to
        $\tau=\SI{0.25}{\second}$, for two different instantaneous opening
        angles $\mean{\theta}$. \textbf{b)} The rotational diffusivity is
        highest for the most compact shapes. \textbf{c)} The joint flexibility
        increases as function of opening angle $\theta$. \textbf{d)} While
        equal for flexed trimers, the translational diffusivity along the long
        axis ($x$) is higher than along the short axis ($y$). \textbf{e)} We
        find a correlation between counter-clockwise rotation and positive $x$
        displacements. \textbf{f)} There is a coupling between translational
        diffusion in the $y$-direction and shape changes: as the cluster
        diffuses in the positive $y$-direction, the angle $\theta$ increases,
    leading to a Brownian scallop-like motion at short timescales. In
panels b-f, the scatter points show the experimental measurements and the lines
show the numerical calculations based on \cite{Harvey1983}.\label{fig:fig2}} 

\end{figure*}

Using \autoref{eq:dtensor}, the resulting shape and time dependent
translational diffusivity in the $y$-direction of twelve rigid and one flexible
trimer are shown in \autoref{fig:fig1}d). Initially, at short timescales, there
is a clear effect of cluster shape for both flexible and rigid trimers:
translational diffusion in $y$ is highest for compact shapes. In comparison to
rigid trimers, the diffusivity of the flexible trimer is slightly enhanced. Two
other features unique to flexible clusters are that using a measurement of only
one cluster, all possible cluster shapes are sampled and the effect of shape
vanishes on a much shorter timescale compared to the rigid clusters.

To study the diffusivity more carefully, we determined the average short time
diffusion tensor of thirty flexible trimers. As shown in \autoref{fig:fig2}a,
the diffusion tensor elements were obtained by fitting the slope of the mean
squared displacement versus lag time. We find three features that are in line with previous findings for rigid clusters~\cite{Kraft2013} and that give confidence in the used analysis: first, translational diffusivity is higher along the longitudinal $x$ direction compared to the lateral $y$ direction (\autoref{fig:fig2}d). Additionally, the
rotational diffusivity shown in \autoref{fig:fig2}b) is higher for compact
trimers as opposed to fully extended trimers and we observe a coupling between
translational diffusion and rotational diffusion in the $x$ direction
(\autoref{fig:fig2}e).

However, flexibility gives rise to other modes that are not present in rigid
assemblies. We found that the flexibility itself, as shown in
\autoref{fig:fig2}c, increases as function of the opening angle, leading to a
four fold increase of flexibility for extended shapes compared to closed
shapes. It is most likely caused by hydrodynamic interactions between the outer
particles, as was predicted by earlier works~\cite{Wegener1982}.

Even more strikingly, the hydrodynamic drag on the outer particles leads to an
increase in opening angle $\theta$ for positive displacements along the $y$
axis (\autoref{fig:fig2}f), which we call the Brownian quasiscallop mode.
We stress that this correlation does not lead to self-propulsion because it has
time reversal symmetry. As the opening angle $\theta$ increases, the location
of the center of mass moves in the negative $y$-direction of the original
particle coordinate system. Therefore, this correlation is larger when the
central particle is chosen as the origin of the coordinate center (see the
Supplementary Information). This Brownian quasiscallop mode may have implications for the accessibility of the functional site in induced fit lock-and-key interactions commonly observed in proteins~\cite{Koshland1995}.

\begin{figure*}[ht]
    \centering
    \includegraphics[scale=1]{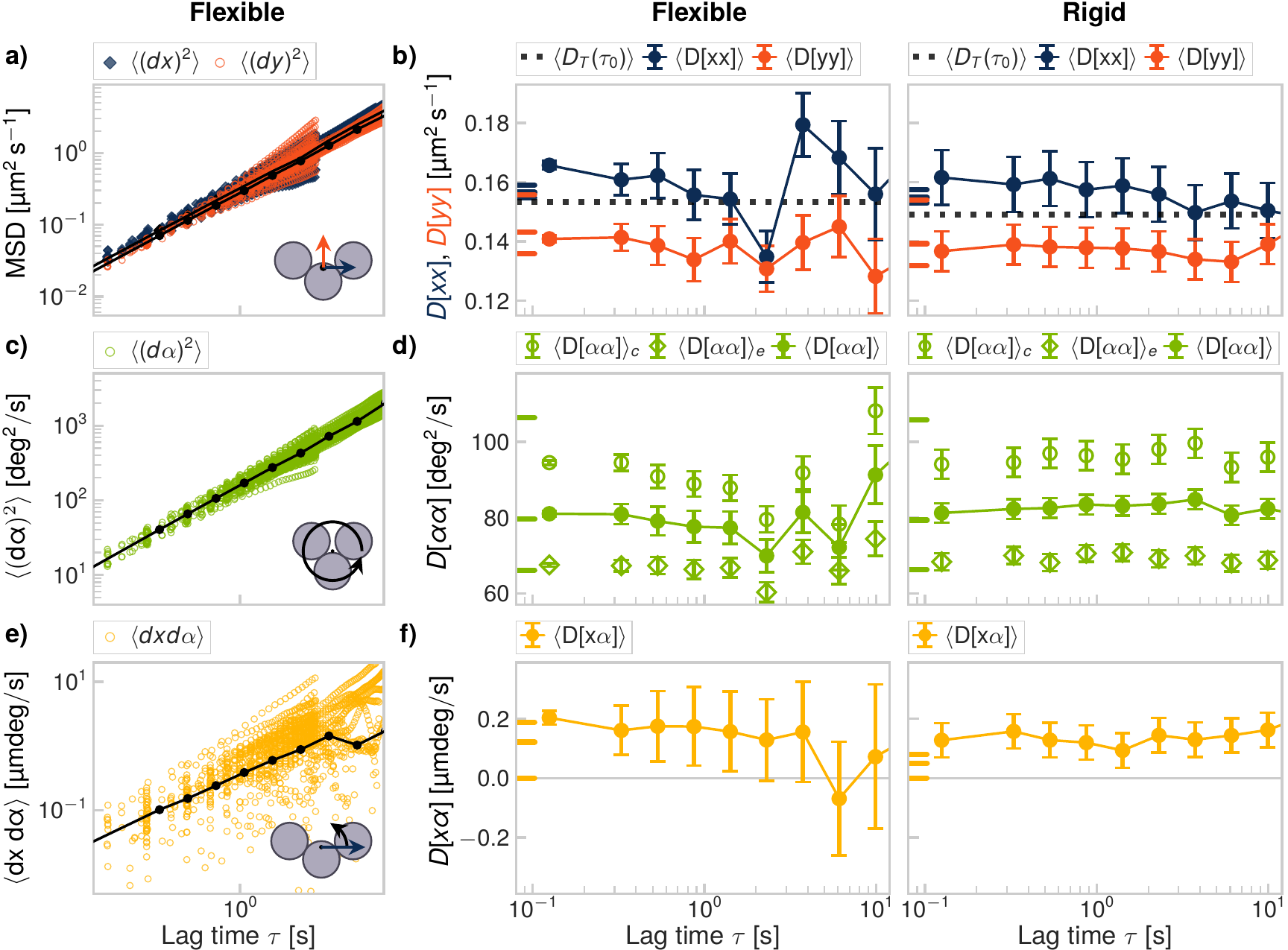}

    \caption{\textbf{Comparison between rigid and flexible trimers.}
        \textbf{a)} Mean squared displacements in $x$ and $y$ for all flexible
        trimers. \textbf{b)} Diffusivity in $x$ and $y$ as function of lag time
        for flexible (left) and rigid (right) trimers. The average translational diffusivity $\langle D_T(\tau_0=\SI{0.25}{\second})\rangle$ (dotted lines) is \SI{2.7\pm0.3}{\percent} higher for flexible clusters compared to rigid clusters. \textbf{c)} Mean squared
        angular displacements for all flexible trimers. \textbf{d)} Rotational
        diffusivity as function of lag time for flexible (left) and rigid
        (right) trimers. $\langle \cdots \rangle_c$ correspond to
        $\mean{\theta}<\SI{120}{\degree}$ (compact) and $\langle \cdots
        \rangle_e$ to $\mean{\theta}\geq\SI{120}{\degree}$ (extended). \textbf{e)} Mean squared coupled displacements in $x$
        and $\alpha$ for all flexible trimers. \textbf{f)} Rotation-translation
        coupling in $x$ and $\alpha$ as function of lag time for flexible
        (left) and rigid (right) trimers. In panels a, c and e,
        colored points are experimental data, black points and
        lines represent the fitted slopes. In panels b, d and f, numerical short-time diffusivities calculated based on \cite{Harvey1983} are indicated by colored ticks on the y-axis, showing minimum, mean, and maximum shape-dependent values from bottom to top.\label{fig:fig3}} 

\end{figure*}

Our experimental data allow us to test for the first time theoretical predictions made by Harvey and coworkers~\cite{Harvey1983}, who modelled the diffusion of segmentally flexible objects by calculating the hydrodynamic interactions between two sub units. We applied their calculations to a bead-shell model, adapted to match the conditions of our experiments (see the Methods and the
Supplementary Information for details) and find good agreement between the
numerical calculations and the experimental data. The good agreement between the numerical results and the experimental data validates their model for the diffusivity of microscopic objects with internal degrees of freedom. For some angles and entries of the diffusion tensor, the experimental data shows small deviations from the predicted model values, especially for translational diffusion, the Brownian quasiscallop mode and the flexibility (see \autoref{fig:fig2}c, d and f). We hypothesize that these differences may
arise because the numerical calculations do not take particle-particle and
particle-substrate friction into account, other than as a first-order
approximate scaling using the friction factors $\phi_{ij}$ as defined in
\autoref{eq:dtensor}. For example, substrate interactions were found to lead to
enhanced diffusion for a model dumbbell consisting of two hydrodynamically
coupled subunits~\cite{Illien2017}. More elaborate models may be used to
provide higher-order corrections to the model we used here~\cite{Swan2007},
however their validity for flexible objects needs to be investigated. Moreover,
our model also does not account for some out-of-plane diffusive motions against
gravity, that might occur in the experiments. Both effects are beyond the scope
of our current work, but we hope they will be investigated in future studies.

Next, we compared the short-term translational, rotational and coupled diffusion coefficients of flexible trimers to rigid trimers that are
frozen in a particular shape and find that while they are qualitatively similar, there are experimentally measurable differences. Specifically, we measure that the average short time diffusion
constant $\langle D_T(\tau_0=\SI{0.25}{\second})\rangle$ of rigid trimers is
\SI{2.7\pm0.3}{\percent} lower (\SI{15\pm2}{\percent} lower without friction
scaling) than that of flexible trimers (\autoref{fig:fig3}a, b, dotted lines),
a small but measurable effect corroborated by the numerical models (see
Supplementary Information). The rotational diffusion constants for flexible
and rigid trimers are equal within the experimental uncertainty (\autoref{fig:fig3}c, d), while the
rotation-translation coupling mode between $x$ and $\alpha$ is slightly higher
for flexible trimers at the shortest lag time (\autoref{fig:fig3}e, f). These
findings agree qualitatively with numerical predictions~\cite{Nagasaka1985,Fixman1981,Akcasu1982} for hinged chains of
spheres of higher aspect ratio (20:1 instead of 3:1 for the trimers). For these
hinged rods, a \SI{10}{\percent} increase in the translational diffusivity and
a higher rotational diffusivity were found compared to rigid rods, which was
attributed to hydrodynamic interactions between the sub-units~\cite{Mellado1988, Harvey1979}.

The last way in which flexibility affects the diffusivity of a cluster is
through the timescales on which effects of the initial cluster shape and orientation on the diffusive motions
vanish. For rigid elongated particles it was shown that the timescale on which translational diffusivity in the $x$ and $y$ directions become
equal with respect to the lab frame is set
by the rotational diffusion time $\gamma_r=(D[\alpha \alpha])^{-1}$, with
$D[\alpha \alpha]$ in \si{rad^2\per\second} \cite{Han2006}. To study this effect for our rigid and freely-jointed trimers, we analyze the
motion of the clusters by defining the lab frame in such a way that the center
of mass of the trimer at lag time $\tau=0$ is at the origin and the
body-centered $x$ and $y$ axes coincide with the original lab frame (see
\autoref{fig:fig1}c), an approach inspired by earlier works on rigid
anisotropic particles~\cite{Chakrabarty2014}. Using the values for the short
time rotational diffusion coefficients for compact and extended trimers, we
find that for both rigid and flexible trimers $\SI{30}{\second} \leq \gamma_{r}
\leq \SI{60}{\second}$. Indeed, by looking at the translational
(\autoref{fig:fig3}b) diffusivity of rigid trimers, we see that the effect of
shape on the diffusivity is preserved up to the maximum lag time we consider
(\SI{10}{\second}). The rotational diffusivity (\autoref{fig:fig3}d) of the rigid trimers stays constant within error (up to at least \SI{10}{\second}).

\begin{figure}[ht]
    \centering
    \includegraphics[scale=1]{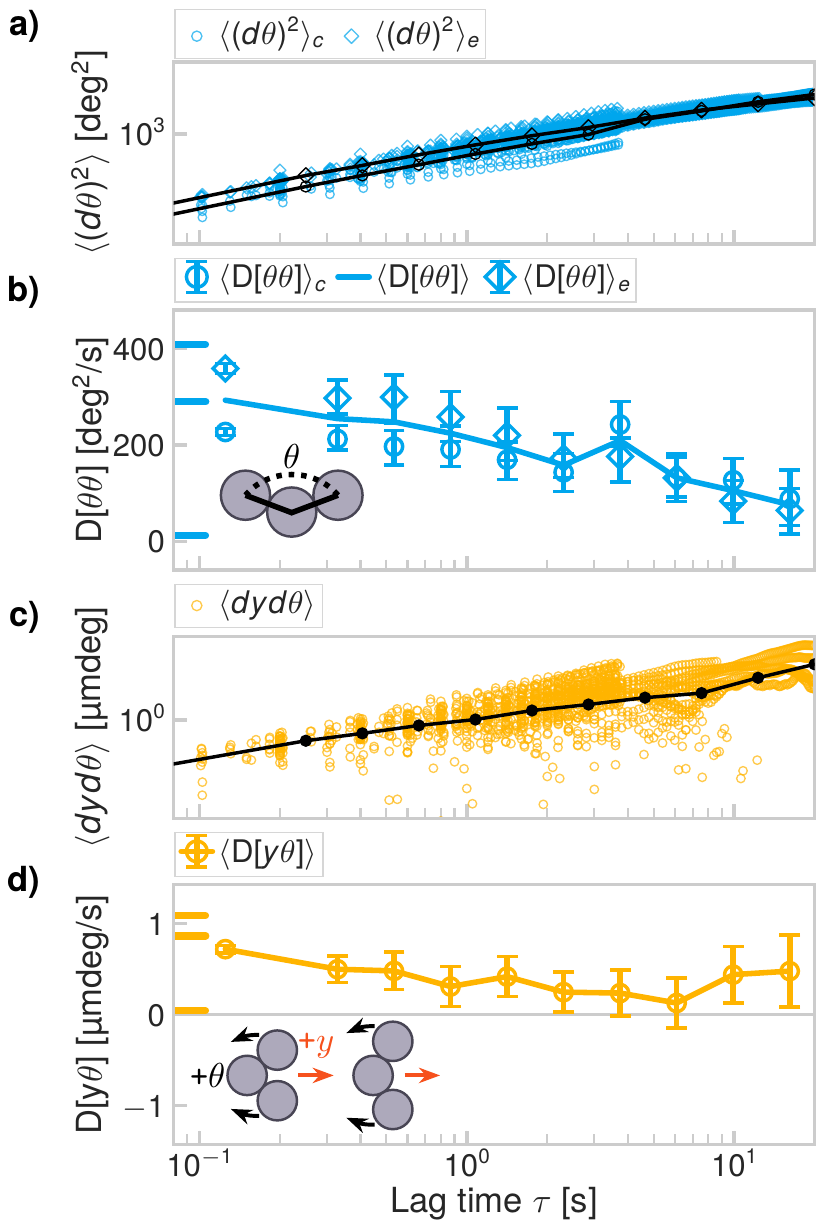}

    \caption{\textbf{Cluster flexibility and Brownian quasiscallop mode as
        function of time.} \textbf{a)} Mean squared angular displacements of
        $\theta$ for all flexible trimers. \textbf{b)} The flexibility
        decreases as function of lag time because of hard-sphere repulsion
        between the two outer particles. \textbf{c)} Mean squared coupled
        displacements of $y$ and $\theta$ for all flexible trimers. \textbf{d)}
        The Brownian quasiscallop mode relaxes on a timescale of a few seconds
        because of conformational and rotational diffusion. In panels a and b,
        $\langle \cdots \rangle_c$ correspond to
        $\mean{\theta}<\SI{120}{\degree}$ (compact) and $\langle \cdots
        \rangle_e$ to $\mean{\theta}\geq\SI{120}{\degree}$ (extended). In
        panels a and c, colored points are experimental data, black
        points and lines represent the fitted slopes. In panels b and d, numerical short time diffusivities calculated based on \cite{Harvey1983} are indicated by colored ticks on the y-axis, showing minimum, mean, and maximum shape-dependent values from bottom to top.\label{fig:fig4}} 

\end{figure}

However, for flexible trimers, the story is different. There exists a second timescale that can average out orientation-dependent effects in diffusion: the timescale of shape changes, which we define as
$\gamma_s=(D[\theta \theta])^{-1}$, analogous to the definition of the
rotational diffusion time. Using the values for the short time flexibility
coefficients for compact and extended trimers, we find that for our flexible
trimers $\SI{8}{\second} \leq \gamma_{s} \leq \SI{35}{\second}$. Therefore, we
hypothesize that for flexible trimers, internal deformations lead to faster
relaxation of the shape-dependency we observe at short lag times and therefore
also the relaxation of differences between translational diffusion in the $x$
and $y$ directions.

Consistent with our hypothesis, the effect of the
initial opening angle appears to be lost on a  shorter timescale than what one would
expect from the rotational diffusion time. In \autoref{fig:fig3}d, the
rotational diffusivity of flexible trimers is not constant in time, as is the
case for rigid trimers, which shows that shape changes affect the diffusivity
at longer lag times. The same effect can be seen in \autoref{fig:fig4}b, where
the cluster flexibility of compact and extended clusters become equal after
about a second. Therefore, for lag times longer than \SI{0.5}{\second}, we only
consider the shape-averaged diffusivities. As can be seen from the
translational diffusivity (\autoref{fig:fig3}b), the shape-averaged diffusivity
in $x$ and $y$ become equal after \SIrange{1}{3}{\second} and this is also the
timescale on which the rotational diffusivity is no longer constant
(\autoref{fig:fig3}d) and the translation-rotation coupling vanishes
(\autoref{fig:fig3}f). Moreover, we observe for both translational, rotational and translation-rotation coupled diffusion that after lag times larger than \SI{2}{\second}, larger fluctuations occur which we attribute to the effect of continuous shape-changes (see \autoref{fig:fig3}b, d and f).

Short timescale relaxation of differences between clusters in extended and
compact conformations exist also for the conformational diffusion tensor
elements. The flexibility (shown in \autoref{fig:fig4}a, b) is smaller for trimers
in flexed conformations than in extended conformations and the difference
vanishes after approximately \SI{2}{\second} due to shape changes.
\autoref{fig:fig4}b shows an overall decrease of flexibility with
lag time, because the range wherein the joint angle can vary is bounded by the two outermost particles.
Furthermore, the magnitude of $D[y\theta]$ (shown in \autoref{fig:fig4}c, d),
which represents the Brownian quasiscallop mode, vanishes on the same timescale
of approximately \SI{2}{\second}, set by the conformational relaxation time
$\SI{8}{\second} \leq \gamma_{s} \leq \SI{35}{\second}$.

\section{Conclusions}

In conclusion, we studied the Brownian motion of flexible trimers and found
features that are unique to flexible objects. We found a hydrodynamic coupling between conformational changes and translations perpendicular to the particle's long axis ($y$-direction), which we call the Brownian quasiscallop mode because of
its resemblance to scallop propulsion at high Reynolds numbers. We found that
this coupling persists over several seconds, a timescale relevant for
biomolecular interactions, implying that it might affect the association of
flexible proteins and other biomolecules. Secondly, we found that the long-time
translational diffusion of the freely jointed trimers was three to fifteen per
cent higher than that of their rigid counterparts. This enhancement was predicted for hinged rods~\cite{Nagasaka1985,Fixman1981,Akcasu1982}, but contrasts with theoretical results on dumbbells of two hydrodynamically coupled subunits, in which extensile shape fluctuations were shown to decrease the translational diffusion coefficient~\cite{Adeleke2019, Illien2017}. Further theoretical and experimental studies are needed to predict the effect of flexibility on diffusivity, since different internal degrees of freedom can have opposing effects.
Finally, we showed that the transition from short- to long-time diffusion
depends not (only) on the rotational diffusion time but mainly on a timescale
related to conformational changes of the particle. We were able to describe our
experimental findings using a hydrodynamic modelling procedure that combines
bead-shell modelling with the approach of Harvey and coworkers
\cite{Harvey1983}. We hope this work inspires other researchers to more
confidently apply this method in the context of the diffusion of segmentally
flexible systems such as biopolymers and proteins.

\section*{Acknowledgements}

This project has received funding from the European Research Council (ERC)
under the European Union's Horizon 2020 research and innovation program (grant
agreement no. 758383) and from the NWO graduate programme.

\section*{Author contributions}

RWV and PM contributed equally to the work. PM, RWV, NL, LPPH performed the
experiments. PM and RWV analysed the data. RWV performed the hydrodynamic
modelling. PM, RWV, JG, WKK, AvB and DJK conceived of the experiments. PM, RWV,
WKK, AvB and DJK wrote the manuscript. All authors discussed the results and
contributed to the final article.

\end{document}


\maketitle
\appendix

\tableofcontents
\listoffigures
\listoftables
 
\clearpage

\section{Modelling and analysis of the hydrodynamic properties of flexible trimers}

\begin{figure}[b]
    \centering
    \includegraphics[width=\textwidth]{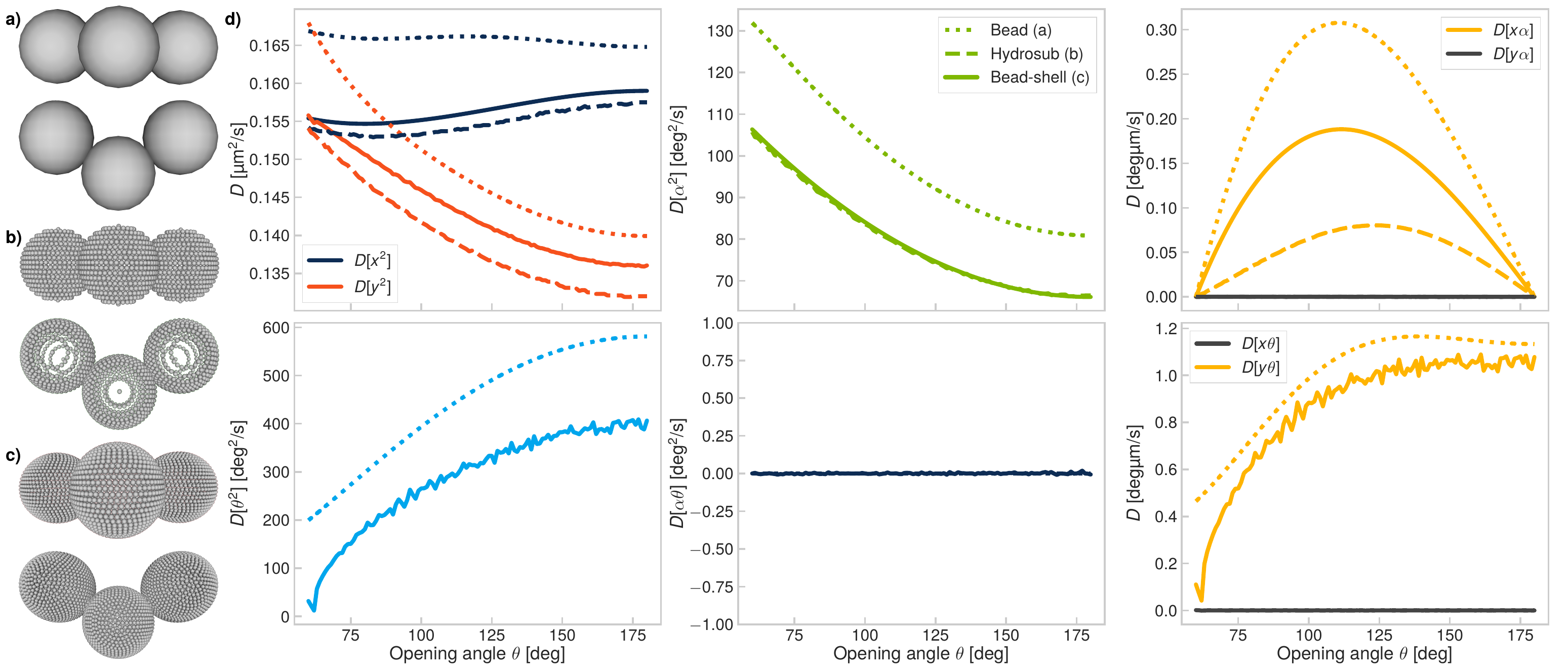}
    \caption[Comparison of the diffusion tensor calculated by different hydrodynamic models]{\textbf{Comparison of the diffusion tensor calculated by different hydrodynamic models.} Renderings made using FreeWRL \cite{FreeWRL} of \textbf{a)} the simple bead model, \textbf{b)} the bead-shell model (minimum radius of the small spheres $r=\SI{55}{\nm}$) used by HydroSub \cite{HydroSub} for rigid trimers, \textbf{c)} the bead-shell model (radius of the small spheres $r=$~\SIrange{31}{54}{\nm}, $r=\SI{45}{\nm}$ is shown) we used for calculating hydrodynamic properties of flexible trimers. For all models, the radius of the large particles is $R=\SI{1}{\um}$. \textbf{d)} Top row, left to right: the translational diffusivity, rotational diffusivity and coupling between translational and rotational diffusivity for the bead model (a, dotted lines), the rigid bead-shell model generated with HydroSub (b, dashed lines) and the segmentally flexible bead-shell model (c, solid lines). Bottom row, left to right: the joint flexibility, coupling between shape changes and rotation and couplings between shape changes and translational diffusion.\label{fig:hydro_models}} 
\end{figure}

\begin{figure}
    \centering
    \includegraphics[width=\textwidth]{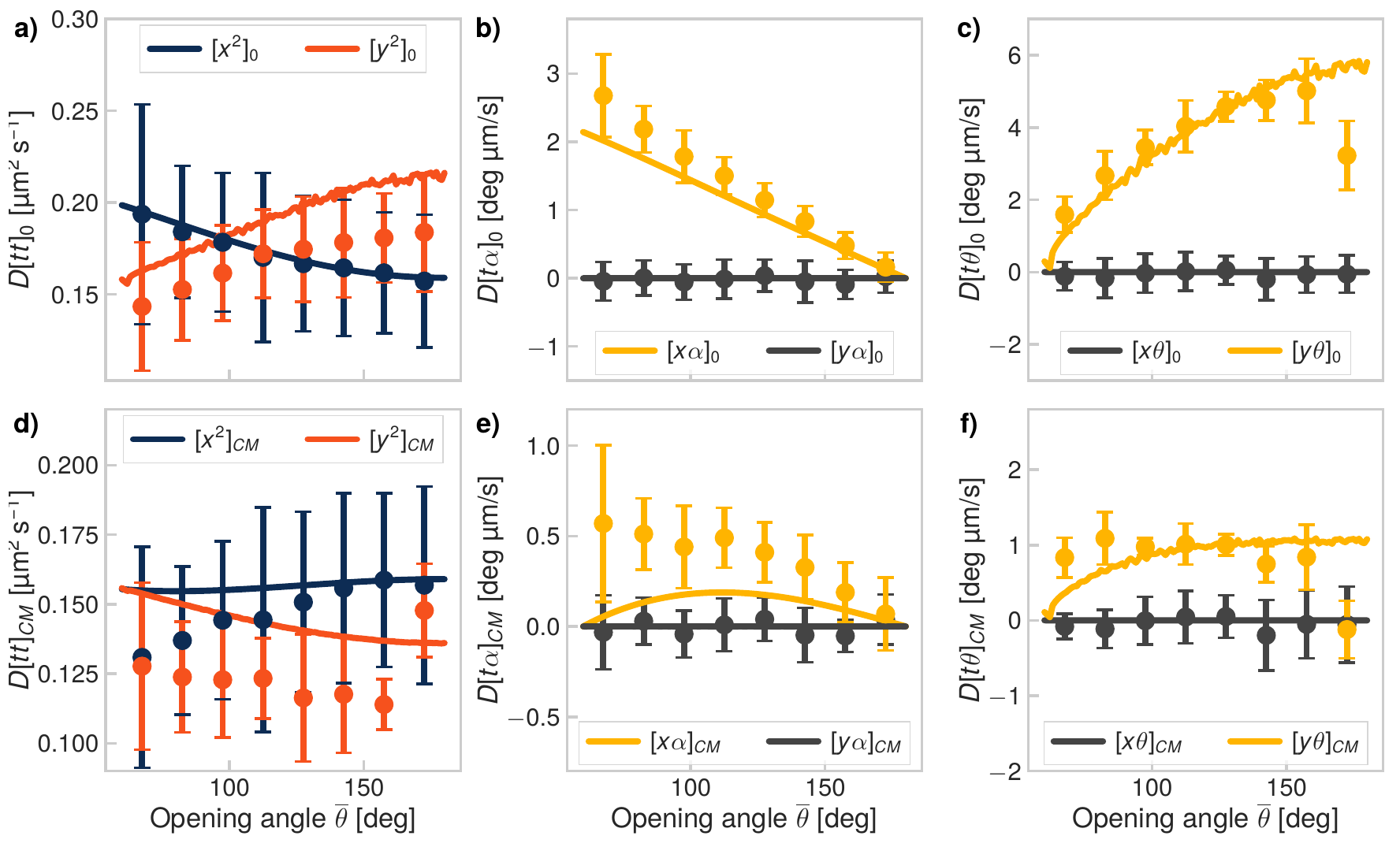}
    \caption[Influence of the reference point on the diffusion tensor]{\textbf{Influence of the reference point on the diffusion tensor.} \textbf{a-c)} The translational (a), translational-rotational (b) and translational-conformational (c) diffusivities with the reference point chosen in the center of the central particle. \textbf{d-f)} The translational (d), translational-rotational (e) and translational-conformational (f) diffusivities with the reference point at the center of mass of the cluster. For these graphs, we transformed the data from panels a-c using the coordinate transformation described in the text from the ``center particle''-based to the ``center of mass''-based diffusivity. Note that the combination of experimental errors of the $D[t\alpha]$, $D[\alpha^2]$, $D[t\theta]$, $D[\theta^2]$ and $D[\alpha\theta]$ terms lead to large uncertainties and deviations, especially for the translational diffusivities. In all panels, the points show the experimental data and the lines are the predictions of the bead-shell model.\label{fig:tracking_point}} 
\end{figure}

We used three different models to describe the hydrodynamic properties of the
flexible trimers: a bead model (\autoref{fig:hydro_models} a), a bead-shell
model for a rigid trimer using HydroSub \cite{HydroSub}
(\autoref{fig:hydro_models} b) and a bead-shell model for flexible trimers
(\autoref{fig:hydro_models} c). For the bead-shell models, the results were
evaluated for multiple small bead sizes, so that the result could be linearly
extrapolated \cite{Bloomfield1967,Filson1967} to the limit where the small bead
radius approaches zero (see the Methods in the main text for details). 

In \autoref{fig:hydro_models} d, the calculated diffusivities are shown for all
three models. The bead model predicts higher diffusivities compared to both
bead-shell models for all different elements of the diffusion tensor. The
bead-shell models agree qualitatively, but predict different magnitudes of the
diffusivities due to differences in hydrodynamic interactions between the outer
beads, which are higher for the flexibly linked clusters
\cite{Mellado1988,Nagasaka1985,Fixman1981,Akcasu1982}. We have used the
bead-shell model of \autoref{fig:hydro_models} c (solid line) to model our
experimental data, because it best describes our experimental data and because
it can be used to model conformational changes, which are not yet implemented in
the HydroSub model.

For our results described in the main text, we used the center of mass as
reference point. For purely rotational and conformational terms, the
diffusivity is expected to be independent of the chosen reference point,
however, for terms that include translation, the location of the reference
point has a large effect on the measured diffusivity
\cite{Harvey1983,Wegener1985}. This can be seen in
\autoref{fig:tracking_point}: in panels a-c), we show the diffusivities
calculated using the central particle as reference point. The results are
remarkably different from the center of mass based results shown in
\autoref{fig:tracking_point} d-f), where we have used the diffusivities
relative to the central particle to calculate the diffusivities relative to the
center of mass using the coordinate transformations determined by Harvey and
coworkers:\cite{Harvey1983} 
\begin{align}
    D[tt]_{CM} &= D[tt]_0 + D[t\alpha]_0^{\intercal} \cdot U - U \cdot D[t\alpha]_0 + U \cdot D[\alpha^2] \cdot U + D[t\theta]_0^{\intercal} \cdot W + W^{\intercal} \cdot D[t\theta]_0 \nonumber \\
               &\phantom{{}=D[tt]_0} - U \cdot D[\alpha\theta]^{\intercal} \cdot W + W^{\intercal} \cdot D[\alpha\theta] \cdot U + W^{\intercal} \cdot D[\theta^2] \cdot W \label{eq:coord_transform_tt}\\
   D[t\alpha]_{CM} &= D[t\alpha]_{0} + D[\alpha^2] \cdot U + D[\alpha\theta]^{\intercal} \cdot W \label{eq:coord_transform_bb} \\
   D[t\theta]_{CM} &= D[t\theta]_{0} + D[\alpha\theta] \cdot U + D[\theta^2] \cdot W \label{eq:coord_transform_aa} 
\end{align}
We have made this comparison because the coupling terms are expected to be
larger in the central particle frame. The results indeed show this larger
coupling and exclude the possibility that the coupling modes we observed are
artifacts of the coordinate system we used. Because the rotational and
conformational diffusivities are independent of the reference point, localization
uncertainties in the determination of the position of the reference point may
have a larger effect on $D[xx,yy,xy,x\alpha,y\alpha,x\theta,y\theta]$ than on
$D[\alpha^2,\theta^2,\alpha\theta]$. Because of the uncertainties that are
propagated when we first determine the diffusivity with respect to the central
particle and then transform this to the diffusivity with respect to the center
of mass (in \autoref{fig:tracking_point} d-f), the error is larger for this
method compared to the direct calculation of the diffusivities with respect to
the center of mass.

\clearpage

\section{Near-wall diffusion: friction factors}

\begin{figure}[hb]
    \centering
    \includegraphics[scale=1]{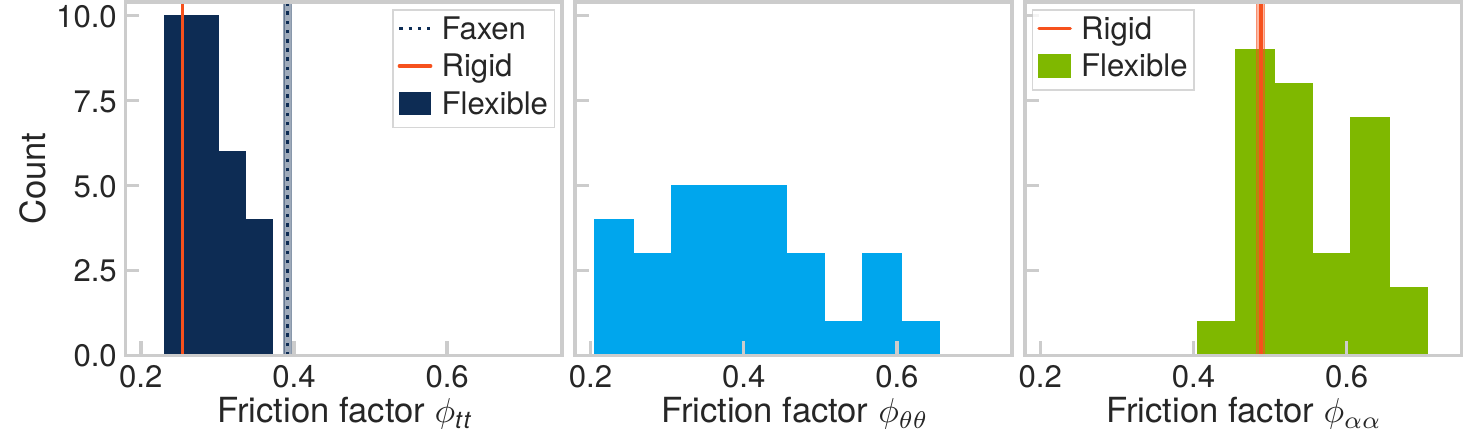}

    \caption[Distribution of friction factors]{\textbf{Distribution of friction factors} as given by equation (3)
        of the main text. The mean value for $\phi_{tt} = 0.29 \pm 0.04$ is
        close to the lower bound of \num{0.39} (indicated by the dotted line)
        as predicted by \autoref{eq:faxen}. We find an average rotational
        friction factor $\phi_{\alpha\alpha} = 0.55 \pm 0.07$. The average
    flexibility friction factor $\phi_{\theta\theta} = 0.40 \pm 0.12$ shows a
broader distribution, which we attribute to a spread in DNA linker
concentration. The average friction factor of the rigid clusters is also
indicated ($\phi_{tt,r} = 0.254 \pm 0.004$, $\phi_{\alpha\alpha,r} = 0.49 \pm 0.02$) and coincides with the friction factors we find for flexible
clusters.\label{fig:phi_dist}} 

\end{figure}

As a first approximation to compare the experimental diffusion of
freely-jointed trimers above a substrate to models of trimers diffusing in the
bulk, we use Faxen's theorem: \cite{Faxen1923}
\begin{align}
    \frac{D_w(h)}{D_0} &= 1 - \frac{9}{16} \frac{R}{h+R} + \frac{1}{8} \left(\frac{R}{h+R}\right)^3 - \frac{45}{256} \left(\frac{R}{h+R}\right)^4 + O\left(\left(\frac{R}{h+R}\right)^5\right), \label{eq:faxen}
\end{align}
with $D_0$ the translational diffusion coefficient in the bulk, $D_w(h)$ the
in-plane translational diffusion coefficient near a wall at height $h$ and $R$
the particle radius.  We calculate an effective particle radius $R_{eff} =
\frac{k_B T}{6 \pi \eta D} =$~\SI{1.8}{\um} from the short-time
translational diffusion coefficient \cite{Ortega2007},
with $k_B$ Boltzmann's constant, $T$ the temperature, $\eta$ the viscosity of
the medium and $D =$~\SI{0.136}{\um\squared\per\second} the lowest short-time
translational diffusion coefficent of the trimer as predicted by the
bead-shell model.

The expected Debye length of our medium (at $I=\SI{200}{\milli M}$) is
$\kappa^{-1} = \frac{0.304}{\sqrt{I}} \approx \SI{0.7}{\nm}$
\cite{Israelachvili1985}, and so we neglect electrostatic interactions between
the trimer and substrate. Therefore, the height of the particle above the
substrate is set by balancing the effect of sedimentation and thermal
fluctuations as expressed by the gravitational length $l_g = \frac{k_B T}{g
\Delta\rho V}$, with $g$ the gravitational acceleration constant, $\Delta\rho$
the density difference between the particle and the medium and $V$ the volume
of the particle.

Using the appropriate values for the trimer, we find $l_g
=$~\SI{20}{\nm}. By setting this as input for $h$ in
\autoref{eq:faxen}, we obtain a upper bound for $\frac{D_w(h)}{D_0}$, equal to
\num{0.40}. A lower bound is found at $h=0$, which gives a value of \num{0.39}.
The translational friction coefficient $\frac{D_w(h)}{D_0}$ that we find has an
average value of \num{0.29\pm0.04}, as shown in \autoref{fig:phi_dist}, which
is close to the lower bound we have calculated above. The experimental value is
slightly lower than the predicted lower bound, because \autoref{eq:faxen}
accounts for hydrodynamic interactions only and real experiments typically show
lower diffusivities because of additional sources of friction
\cite{Sharma2008}, which in the present case could be explained by additional
friction between the polymer coating and the particles.

\clearpage

\section{The collective diffusion constant depends on size polydispersity}

In this manuscript we showed that freely-jointed trimers diffuse slightly faster than rigid trimers; their diffusion constant differs by $3\%$. When reporting such a small difference it is important to exclude other effects that could lead to similar variations in the diffusion constant. Therefore, in this section we address the effect of size polydispersity on the average diffusion constant of a collection of particles. We consider an ensemble of particles, whose sizes are normally distributed around an average radius, $\bar{a}$, and with a standard deviation, $\sigma$. We assume that the particles exhibit Stokes diffusion so that each particle $i$ has a size dependent diffusion constant $D_i=\frac{k_B T}{6 \pi \eta a_i}$. The small particles in the ensemble diffuse faster than large particles.

The experimental average diffusion constant of this ensemble of particles, $\bar{D}$, can be found by tracking the motion of many individual particles, calculating their individual diffusion constants and averaging those. One might assume that this average diffusion constant equals the diffusion constant of a monodisperse sample of particles with the same average size, but this turns out to be generally not true: $\bar{D} \neq \frac{k_B T}{6 \pi \eta \bar{a}}\equiv D_{\bar{a}}$. The reason for this inequality is that the diffusion constant scales nonlinearly with size. Therefore, the diffusion constants of small particles are weighted more heavily than those of large particles, which skews the distribution of diffusion constant and shifts the average away from $D_{\bar{a}}$.

\begin{figure}[hb]
    \centering
    \includegraphics[scale=1]{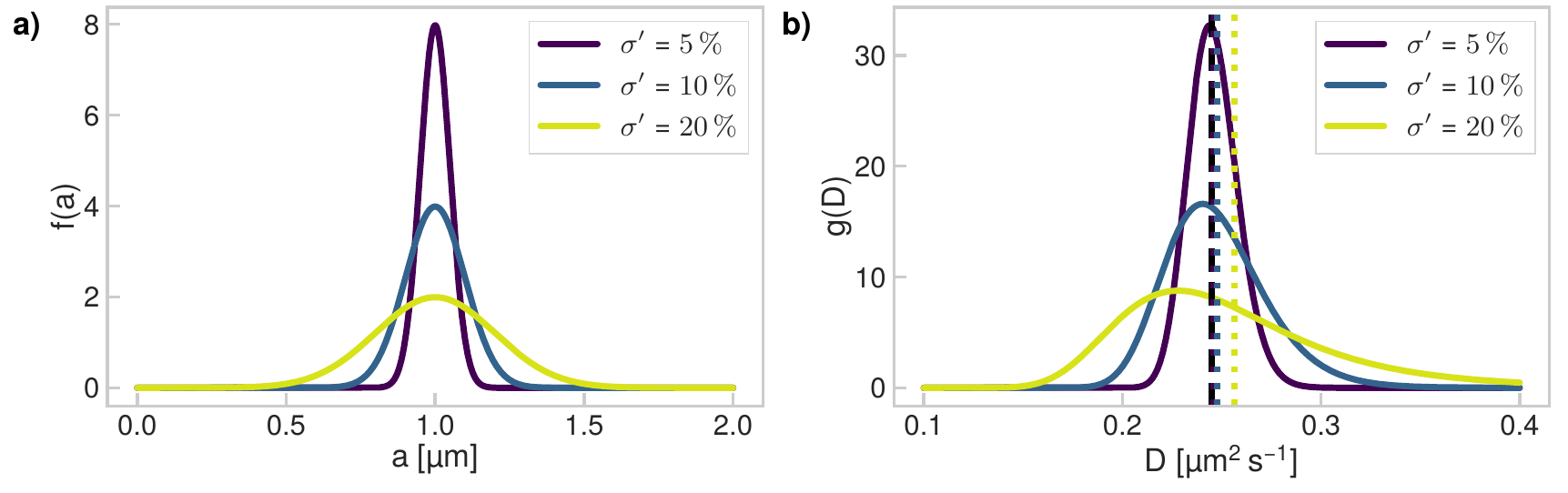}
    \caption[The collective diffusion constant depends on size polydispersity]{\textbf{The collective diffusion constant depends on size polydispersity.} \label{fig:distributions} \textbf{a)} Three hypothetical particle size distribution with an average particle radius of $2~\mu m$ and relative polydispersities of $5\%$, $10\%$, and $20\%$. \textbf{b)} The distributions in diffusion constant corresponding to the three particle size distribution in panel a. The average diffusion constants are indicated by dashed lines. The average diffusion constant of a monodisperse sample is indicated by a black dashed line.}
\end{figure}

We asked how much the collective diffusion constant of a polydisperse sample would deviate from that of a monodisperse sample and how this deviation depends on size polydispersity. To this end, we first define the relative polydispersity as $\sigma'=\sigma/\bar{a}$, which is a value between 0 and 1. The normalized distribution of particle sizes is then
\begin{equation}
\label{PDFa}
f(a)=\frac{1}{\sigma' \bar{a} \sqrt{2 \pi}} \exp\left[-\frac{1}{2}\left(\frac{a-\bar{a}}{\sigma'\bar{a}}\right)^2\right].
\end{equation}
Because the size is normally distributed and the diffusion constant scales with size as $1/a$, the diffusion constant exhibits a reciprocal normal distribution:
\begin{equation}
\label{PDFD}
g(D)=\frac{D_{\bar{a}}}{D^2 \sigma'  \sqrt{2 \pi}} \exp\left[-\frac{1}{2}\left(\frac{D_{\bar{a}}}{\sigma'}(1/D-1/D_{\bar{a}})\right)^2\right].
\end{equation}
\autoref{fig:distributions}a shows the hypothetical size distributions of three sets of particles with an average radius of $1~\mu m$ and relative polydispersities of  $5\%$, $10\%$, and $20\%$. \autoref{fig:distributions}b shows the diffusion constant distributions that correspond to these particle ensembles. Note that the diffusion constant is --- unlike the size --- not normally distributed, but has a tail of fast diffusion, corresponding to small particle sizes. Note also that the most probable diffusion constant  shifts with polydispersity. This is also due to the $1/a$ scaling of the diffusion constant and can intuitively be explained by the fact that a range of large particles give a similarly small diffusion constant. This increases the probability of measuring this small diffusion constant and shifts the peak in the distribution. These properties of the distribution cause the average diffusion constant of a polydisperse sample (indicated by dashed lines in \autoref{fig:distributions}b) to shift compared to the monodisperse case (indicated by a black dashed line). How much the diffusion constant is underestimated depends on the size polydispersity. We intentionally chose large polydispersities to show the effect clearly. Note that at for a size polydispersity of $5\%$ the distribution of diffusion constants still looks rather symmetric.

The average diffusion constant of the particle ensemble is
\begin{equation}
\label{Dav}
\bar{D}=\int_{D=-\infty}^{D=\infty} D~g(D)~dD.
\end{equation}
The integral in Equation~\ref{Dav} cannot be solved analytically, but we solved it numerically and compared it to the diffusion constant corresponding to particles with an average size $D_{\bar{a}}$. As integration limits we used $0$ and $100\times D_{\bar{a}}$ in order to probe all non-zero elements of the distribution function. We found that a $5\%$ polydispersity results in an underestimate of the diffusion constant by only $0.25\%$. To underestimate the diffusion constant by $3\%$, the relative polydispersity needs to be at least $17\%$. We found that these results are independent of the particle size. This finding indicates that the measured $3\%$ increase of flexible trimers compared to rigid trimers cannot be due to size polydispersity alone, because the employed particles have a size polydispersity of only \SI{2.6}{\percent}.

While polydispersity does not drastically alter the collective diffusion of microparticle suspensions, where $\sigma'$ is typically around $5\%$, it could play a large role in the diffusion of nanoparticles, where a $\sigma'$ on the order of $100\%$ is not uncommon \cite{Weare2000}. For example, gold nanoparticles with relative polydispersties on order of $10\%$ are considered very monodisperse and can only be made in a small parameter range\cite{Piella2016}. Using Equation~\ref{Dav} we predict that the collective diffusion constant of a sample with $100\%$ polydispersity is $63\%$ larger than a monodisperse sample with the same average size, highlighting the importance of considering this effect in nanoparticle suspensions.

\clearpage

\section{Supplementary Figures}

\begin{figure}[hb]
    \centering
    \includegraphics[scale=1]{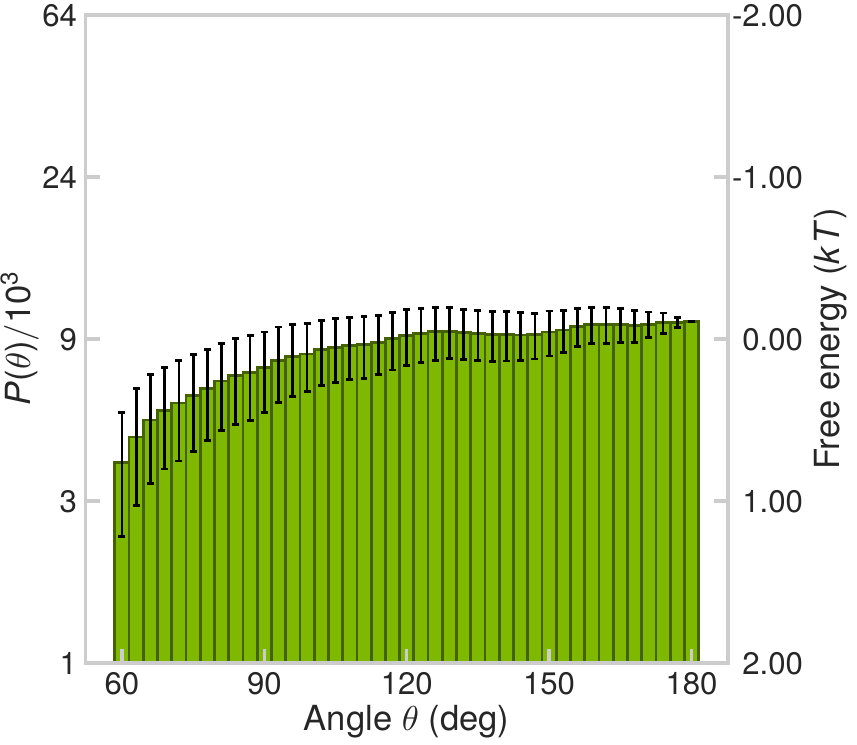}
    \caption[Probability and free energy as a function of the opening angle of flexible trimers]{\textbf{Probability and free energy as a function of the opening angle of flexible trimers.} Probability and corresponding free energy of the opening angle of the flexible trimers used in this work (with the reference set at \SI{180}{\degree}). There is no preference for a specific opening angle within the experimental error, meaning the particles are freely-jointed, as was shown before. \cite{Rinaldin2019} Note that the slightly lower probability at angles smaller than $\SI{60}{\degree}+\sqrt{2J\tau}\approx\SI{69}{\degree}$ (with $J$ the joint flexibility and $\tau$ the sampling interval) is caused by boundary effects inherent to our analysis method. \cite{Rinaldin2019}\label{fig:freely-jointed}} 
\end{figure}

\begin{figure}[hb]
    \centering
    \includegraphics[scale=1]{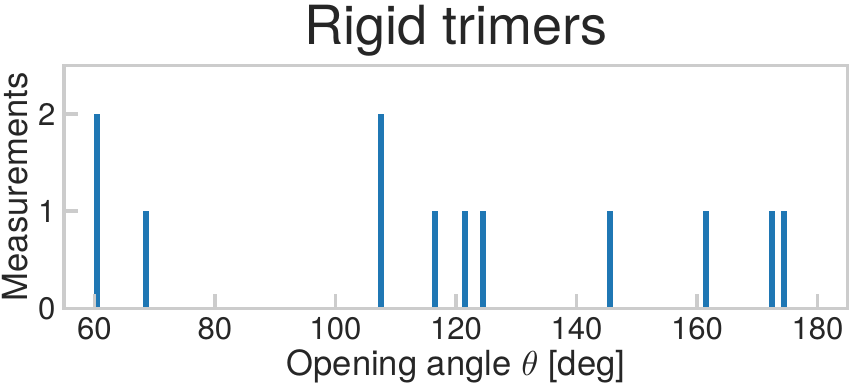}
    \caption[Opening angles of rigid trimers]{\textbf{Opening angles of rigid trimers.} The number of rigid clusters of different opening angles used in this study. Six rigid trimers have a `compact' opening angle (below \SI{120}{\degree}) while the other six are more extended.\label{fig:thetasrigid}} 
\end{figure}

\begin{figure}[hb]
    \centering
    \includegraphics[scale=1]{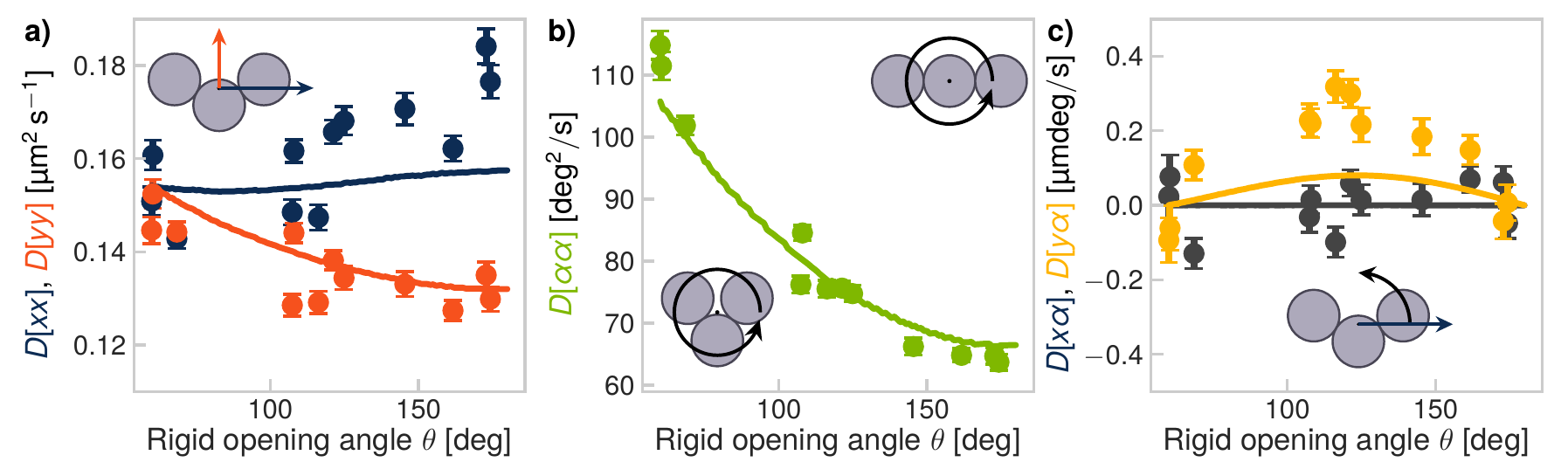}
    \caption[Short-time diffusion of rigid trimers]{\textbf{Short-time diffusion of rigid trimers.} The \textbf{(a)} translational,  \textbf{(b)}  rotational and \textbf{(b)} translational-rotational diffusivities of rigid trimers with various opening angles (see \autoref{fig:thetasrigid}). In all panels, the points correspond to the experimental diffusivities (up to lag times of \SI{0.25}{\second}) and the solid lines correspond to the numerical calculations performed using HydroSub \cite{HydroSub}. All points are scaled by the same average friction factor as shown in \autoref{fig:phi_dist} in order to compare the experiments to the numerical simulations.\label{fig:strigid}} 
\end{figure}

\clearpage

\section{Supplementary Tables}

\begin{table}[hb]
    \caption[Overview of all DNA strands used]{\textbf{Overview of all DNA strands used}. Sticky ends are marked in cursive.\label{table:DNA}}\vspace{1em}
    \centering
    \begin{tabularx}{\textwidth}{lX}
        \toprule
        \textbf{Identifier} & \textbf{Sequence} \\ \midrule
        DS-B & \texttt{5$'$\--TCG\--TAA\--GGC\--AGG\--GCT\--CTC\--TAG\--ACA\--GGG\--CTC\--TCT\--GAA\--TGT\--GAC\--TGT\--GCG\--AAG\--GTG\--ACT\--GTG\--CGA\--AGG\--GTA\--GCG\--ATT\--TT\--3$'$}\\
        DS-S-A & \texttt{Double Stearyl\--HEG\--5$'$\--TT\--TAT\--CGC\--TAC\--CCT\--TCG\--CAC\--AGT\--CAC\--CTT\--CGC\--ACA\--GTC\--ACA\--TTC\--AGA\--GAG\--CCC\--TGT\--CTA\--GAG\--AGC\--CCT\--GCC\--TTA\--CGA\--\textit{GTA\--GAA\--GTA\--GG}\--3$'$\--6FAM}\\
        DS-S-B & \texttt{Double Stearyl\--HEG\--5$'$\--TT\--TAT\--CGC\--TAC\--CCT\--TCG\--CAC\--AGT\--CAC\--CTT\--CGC\--ACA\--GTC\--ACA\--TTC\--AGA\--GAG\--CCC\--TGT\--CTA\--GAG\--AGC\--CCT\--GCC\--TTA\--CGA\--\textit{CCT\--ACT\--TCT\--AC}\--3$'$\--Cy3} \\
        PA-A & \texttt{Cholesterol\--5$'$\--TTT\--ATC\--GCT\--CCC\--TTC\--GCA\--CAG\--TCA\--ATC\--TAG\--AGA\--GCC\--CTG\--CCT\--TAC\--GAT\--ATT\--GTA\--CAA\--TA\--3$'$\--Cy3} \\
        PA-B & \texttt{Cholesterol\--5$'$\--CGT\--AAG\--GCA\--GGG\--CTC\--TCT\--AGA\--TTG\--ACT\--GTG\--CGA\--AGG\--GTA\--GCG\--ATT\--TT\--3$'$} \\
        DS-H-A & Obtained by hybridization of DS-B and DS-S-A \\
        DS-H-B & Obtained by hybridization of DS-B and DS-S-B \\
        PA\--C & Obtained by hybridization of PA\--A and PA\--B \\
        \bottomrule
    \end{tabularx}
\end{table}